\documentclass[aps,pre,amsmath,superscriptaddress,amssymb,twocolumn]{revtex4}     
\usepackage{graphicx}

\newcommand*{\nmax}{n_{\rm max}}

\begin{document}
\title{Free energy of formation of small ice nuclei near the Widom line in simulations of supercooled water}

\author{Connor R. C. Buhariwalla}
\affiliation{Department of Physics, St. Francis Xavier University,
Antigonish, NS, B2G 2W5, Canada}

\author{Richard K. Bowles}
\affiliation{Department of Chemistry, University of Saskatchewan, Saskatoon, SK, 57N 5C9, Canada}

\author{Ivan Saika-Voivod}
\affiliation{Department of Physics and Physical Oceanography, Memorial University of Newfoundland, \\St. John's, Newfoundland A1B 3X7, Canada}

\author{Francesco Sciortino}
\affiliation{Dipartimento di Fisica,
Universit\`a di Roma {\em La Sapienza},
Piazzale A. Moro 5, 00185 Roma, Italy}

\author{Peter H. Poole}
\email{ppoole@stfx.ca}
\affiliation{Department of Physics, St. Francis Xavier University,
Antigonish, NS, B2G 2W5, Canada}

\begin{abstract}
The ST2 interaction potential has been used in a large number of simulation studies to explore the possibility of a liquid-liquid phase transition (LLPT) in supercooled water.  
Using umbrella sampling Monte Carlo simulations of ST2 water, we evaluate the free energy of formation of small ice nuclei in the supercooled liquid in the vicinity of the Widom line, the region above the critical temperature of the LLPT where a number of thermodynamic anomalies occur.  Our results show that in this region there is a substantial free energy cost for the formation of small ice nuclei, demonstrating that the thermodynamic anomalies associated with the Widom line in ST2 water occur in a well-defined metastable liquid phase.  On passing through the Widom line, we identify changes in the free energy to form small ice nuclei that illustrate how the thermodynamic anomalies associated with the LLPT may influence the ice nucleation process.
\end{abstract}

\date{\today}
\maketitle

\section{Introduction}

The proposal that a liquid-liquid phase transition (LLPT) occurs in supercooled water has stimulated a significant quantity of research and debate~\cite{PSES,harr,stanley,stanley2,pdreview,faraone,xu,denmin,liu1,lim,sus, cuth,becker,wink,zhang,kessel,jagla,liu,poole2013,limmer2013,rsc,wink1,sfs2014,palmer2014,holten,dc,palmerarxiv,yaga,starr,starrpnas,starrnat}.
The principal virtue of this proposal is that it provides a common origin for two of water's most unusual properties. First, the atypical and increasingly divergent behaviour of thermodynamic and dynamical properties of liquid water as temperature decreases through the freezing temperature can be understood in terms of an approach to the critical point of the LLPT, and its associated anomalies.  Second, the line of first-order phase transitions that terminates at the critical point of the LLPT can account for the occurrence of two distinct glass forms of water, the so-called low-density amorphous and high-density amorphous ices.  

\begin{figure}\bigskip\bigskip
\centerline{\includegraphics[scale=0.35]{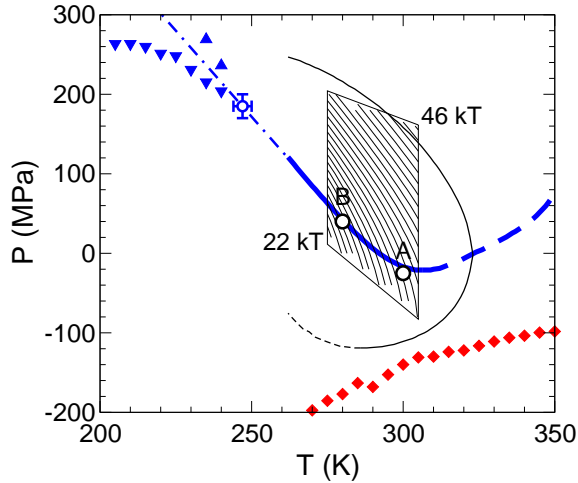}}
\caption{Phase behaviour of the ST2 model predicted from previous MD simulations~\cite{denmin,cuth}.  Shown are the estimated locations of the liquid-liquid critical point (circle with error bars); the HDL spinodal (down-triangles) and the LDL spinodal (up-triangles); the line of $K_T$ maxima (thick solid line) and minima (thick dashed line);  the line of density maxima (thin solid line) and minima (thin dashed line); and the metastability limit of the liquid at negative pressure (diamonds).  The dot-dashed line is a line having the estimated slope of the liquid-liquid coexistence curve at the critical point.  The circles labelled $A$ and $B$ locate the state points at which we carry out the umbrella sampling simulations described in the text.  The enclosed polygon delimits the region of states within which we estimate the free energy of formation of small ice nuclei.  The contours within this polygon are lines along which $G_{15}$ is constant.  The contours are $1kT$ apart, and range from $22kT$ in the lower left corner to $46kT$ in the upper right.}
\label{PD}
\end{figure}

The LLPT proposal predicts the occurrence of an interrelated set of thermodynamic features in supercooled water.  These features are illustrated in the plane of temperature $T$ and pressure $P$ in Fig.~\ref{PD}, which summarizes the behaviour reported in previous molecular dynamics (MD) simulations of the ST2 model of water~\cite{denmin,cuth}.  The coexistence line separating the distinct low-density liquid (LDL) and high-density liquid (HDL) phases has negative Clapeyron slope and terminates at a critical point at $T=T_c$.  Emanating from the critical point for $T<T_c$, and bracketing the coexistence line, are two metastability limits (spinodals), one for each of the LDL and HDL phases.  For $T>T_c$, a number of thermodynamic response functions (e.g. the isothermal compressibility $K_T$) exhibit maxima which diverge as $T\to T_c$.  The loci of these maxima in the $(T,P)$ plane all lie near the so-called Widom line~\cite{xu}, the locus of maxima of the correlation length, and converge at the critical point.  The region near the Widom line therefore contains the high-temperature harbingers of the critical point and the line of first-order phase transitions that occur for $T\leq T_c$.  Whereas the HDL-LDL transition is discontinuous at the line of first-order transitions, the liquid continuously crosses over from an HDL-like liquid to an LDL-like liquid as $T$ decreases through the region of the Widom line.

The main source of challenge and concern related to the LLPT hypothesis is ice formation.  Experimentally, rapid ice formation has so far thwarted the direct measurement of the properties of bulk supercooled liquid water in the region of $T$ and $P$ where the LLPT is predicted to occur~\cite{pdreview}.  That is, neither the first-order LLPT, the critical point, nor the response function maxima associated with the Widom line have been directly observed in experiments on bulk samples of supercooled liquid water.  However, a number of indirect experimental results are consistent with the LLPT proposal.  Examples include experimental studies of metastable ice melting lines~\cite{stanley2}, the behaviour of the amorphous ices~\cite{wink,wink1}, and the thermodynamic and transport properties of confined liquid water~\cite{faraone,zhang}.  Recent experiments continue to push deeper into the supercooled liquid regime, offering new insights~\cite{nilsson}.

To date, it is only in computer simulations of water-like models that a LLPT has been unambiguously observed~\cite{palmer2014,sfs2014}.  The most extensively studied model in this regard is the ST2 interaction potential~\cite{st2}.  In ST2 simulations, most previous work shows that it is possible to evaluate the equilibrium properties of the deeply supercooled liquid on a time scale that is short compared to the crystal nucleation time, thus exposing the physics of the LLPT to direct observation~\cite{yaga,palmer2014}.  However, Limmer and Chandler have disputed this interpretation of the results for ST2 water, and have specifically proposed that the behaviour previously interpreted as evidence of a LLPT is in fact a manifestation of crystal formation~\cite{lim,limmer2013,dc}.  Subsequent studies by others using very similar methods do not reproduce the simulation results of Limmer and Chandler, and confirm the occurrence of a LLPT in ST2 water~\cite{liu,poole2013,rsc,palmer2014}.  The origin of this discrepancy remains a subject of debate~\cite{dc,palmerarxiv}.

In light of the above results, the nature of crystal formation in a liquid exhibiting a LLPT, and in ST2 water in particular, merits further investigation.   The relationship between ice formation and the potential for a LLPT varies widely in water-like computer models.  For example, in the mW model of water, recent work shows that the supercooled liquid becomes unstable to ice formation before the conditions for a LLPT can be realized~\cite{mw,mw1}.  At the other extreme, simulations of a water-like patchy colloidal liquid, and a model of DNA tetramers, demonstrate that a LLPT can occur in the stable liquid region, where crystal formation is impossible~\cite{sfs2014,starrpnas,starr,starrnat}.  The ST2 model provides a case where the physics of a LLPT and of ice nucleation occur simultaneously, allowing the influence of the LLPT on ice formation to be explored directly.  The coincident appearance of both phenomena can be understood in terms of the effective rigidity of hydrogen bonds when viewed against a broader spectrum of bond flexibility in tetrahedral network-forming liquids~\cite{ivanadd,starr,sfs2014}.

In the most recent work on ST2 water, the free energy surface of the system near the critical point of the LLPT was examined as a function of the density $\rho$ and the bulk bond-order parameter $Q_6$, using the method of umbrella sampling~\cite{lim,limmer2013,liu,poole2013,rsc,palmer2014}.   $Q_6$ is a bulk measure of the overall crystallinity of a system developed by Steinhardt et al.; see e.g. the definition of $Q_6$ given in Ref.~\cite{lim}.  As such $Q_6$ is a valid order parameter for distinguishing between the disordered liquid and ordered crystal phases, and provides a reaction coordinate to quantify the variation of the free energy associated with the crystal-liquid phase transition.

We focus here on the related question of how single ice nuclei form and grow from the supercooled liquid.  The free energy $G(n)$ to form an ice-like cluster of $n$ molecules from the liquid can be evaluated from computer simulations using,
\begin{equation}
\beta G(n)=-\ln \frac{{\cal N}(n)}{N},
\label{G}
\end{equation}
where $\beta=1/kT$, $k$ is Boltzmann's constant, $N$ is the number of molecules in the system, and ${\cal N}(n)$ is the equilibrium average of the number of ice-like clusters of size $n$~\cite{frenkel}.  So defined, $G(n)$ can be used to determine the nucleation free energy barrier appropriate for estimating the nucleation rate, e.g. via classical nucleation theory~\cite{valeriani}.  When using umbrella sampling methods to compute ${\cal N}(n)$, recent work has shown that an appropriate order parameter to define a path from the liquid to the crystal phase is $n_{\rm max}$, the size of the largest crystalline cluster in the system~\cite{isv2011,vega2012,russo2014}.  In particular, Ref.~\cite{vega2012} discusses why the order parameter $n_{\rm max}$ is preferable to order parameters such as $Q_6$ for evaluating the free energy of formation of crystal nuclei when the aim is to estimate the nucleation barrier and rate.

In order to fully elucidate the influence of a LLPT on crystal nucleation, it would be valuable to determine $G(n)$ for ST2 water near the critical point of the LLPT, and for a sufficient range of $n$ to find the nucleation free energy barrier and the size of the critical nucleus.  However, in order to achieve this goal, a number of concerns present themselves.  For example, $G(n)$ has not been previously reported for the ST2 model, and has been found for only a few other water-like models when using $n_{\rm max}$ as the order parameter~\cite{isv2011,russo2014}.  An initial assessment of the methods for finding $G(n)$ for ST2 water is therefore warranted.  In addition, relaxation times near the critical point of the LLPT are large, relative to easily accessed computational time scales, and it is therefore appropriate to work first in a regime where equilibration time scales are readily attained.

Consequently, in the present work, we evaluate $G(n)$ in ST2 water for small ice nuclei in the range $0<n<20$, and focus on $T$ and $P$ in the vicinity of the Widom line.  These choices allow us to test the methods for estimating $G(n)$ for ST2 water using a relatively small system ($N=216$ molecules) and in a range of $T$ and $P$ where equilibrium is easily established.  Further, for finding $G(n)$, we explore the use of umbrella sampling with respect to the two order parameters $\rho$ and $n_{\rm max}$.  As described below, umbrella sampling with respect to $\rho$ allows us to reweight our results with respect to $P$ over a range near the chosen $P$ of our simulations.  We also conduct separate series of runs at two values of $T$, and are able to interpolate results in between by reweighting with respect to $T$.  The result is an estimate of $G(n)$ inside a continuous region of states in the $(T,P)$ plane straddling the Widom line.  Thus, although the present study does not explore the immediate vicinity of the critical point of the LLPT, our results allow us to draw conclusions about the stability of the metastable liquid state near the Widom line, and to discern how the thermodynamic anomalies that extend outward from the LLPT may influence ice nucleation.

\section{Model and order parameters}

We study the ST2 model of water proposed by Stillinger and Rahman~\cite{st2}.  The details of our implementation of the ST2 interaction potential are summarized in Ref.~\cite{poole2013}, and are the same as those used in a number of previous studies~\cite{PSES,harr,denmin,cuth,becker,sus}.  In particular, we note that in our implementation of the ST2 model, we use the reaction field method to approximate the long-range contribution of the electrostatic interactions~\cite{rf}.

\begin{figure}
\centerline{\includegraphics[scale=0.32]{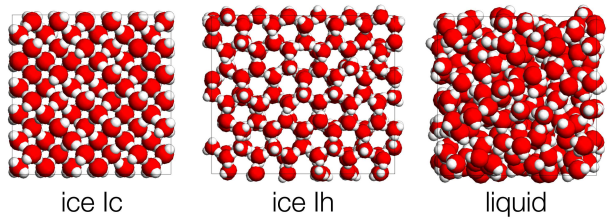}}
\caption{Representative configurations of ice Ic, ice Ih, and liquid water obtained from MC simulations at $T=280$~K and $P=40$~MPa.  The simulations of ice~Ic and the liquid are conducted in a cubic simulation cell with $N=216$ molecules.  The ice~Ih simulation is carried out in a rectangular cell with $N=432$ molecules.}
\label{ices}
\end{figure}

\begin{figure}\bigskip
\centerline{\includegraphics[scale=0.35]{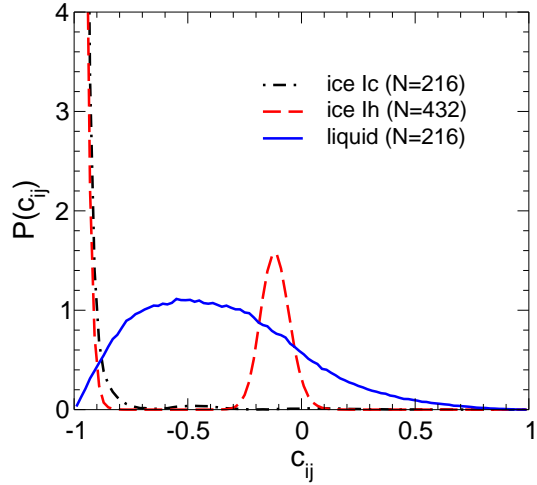}}
\caption{Normalized probability distributions $P(c_{ij})$ representing the frequency of occurrence of $c_{ij}$ values for the equilibrium liquid, ice~Ic, and ice~Ih phases at $T=280$~K and $P=40$~MPa.}
\label{cij}
\end{figure}

\begin{figure}\bigskip
\centerline{\includegraphics[scale=0.35]{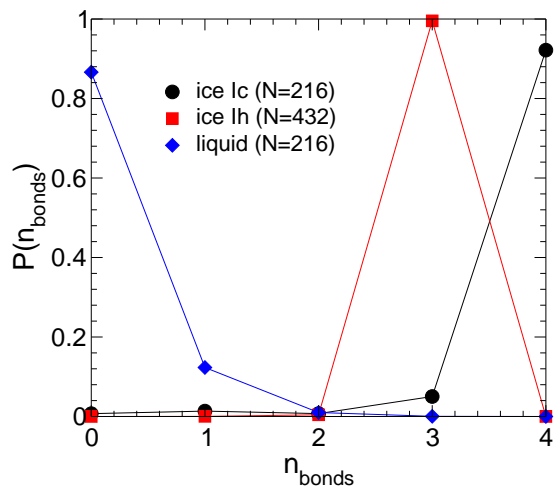}}
\caption{Normalized probability distributions $P(n_{\rm bonds})$ for the number of crystalline bonds $n_{\rm bonds}$ found for molecules in the equilibrium liquid, ice~Ic, and ice~Ih phases at $T=280$~K and $P=40$~MPa.}
\label{bonds}
\end{figure}

To characterize the crystallinity of the system, we follow the approach of Refs.~\cite{rss,isv2011}.
We use Steinhardt bond order parameters~\cite{q6} based on spherical harmonics $Y_{lm}$ of order $l = 3$.  For each particle $i$, and for integers $m$ such that $-l\le m\le l$, we define the components of a complex vector as,
\begin{equation}
q_{lm}(i)=\frac{1}{4} \sum_{k=1}^{4}Y_{lm}({\hat r}_{ik}),
\end{equation}
where the sum is over the four nearest neighbours of molecule $i$, and ${\hat r}_{ik}$ is a unit vector pointing from the O atom of molecule $i$ to that of molecule $k$.  The dot product,
\begin{equation}
c_{ij}=\sum_{m=-l}^{l}{\hat q}_{lm}(i)\,{\hat q}^\ast_{lm}(j),
\end{equation}
where,
\begin{equation}
{\hat q}_{lm}(i)=\frac{q_{lm}(i)}{\sqrt{\sum_{m=-l}^{l} |q_{lm}(i)|^2}},
\end{equation}
and ${\hat q}^\ast_{lm}(i)$ is its complex conjugate, quantifies the orientational correlation between the nearest-neighbour shells of molecules $i$ and $j$. 

We carry out initial simulations of ST2 water in the liquid phase, and in the ice~Ic and Ih crystalline phases, all at the same state point at $T=280$~K and $P=40$~MPa; see Fig.~\ref{ices}.  Fig.~\ref{cij} shows the normalized probability distributions of $c_{ij}$ for each phase.  The distribution for the liquid phase is broad and spread out over much of the range of $c_{ij}$.  The ice~Ic distribution is peaked sharply around $c_{ij} =-1$, reflecting the fact that each molecule participates in four ``eclipsed" hydrogen bonds in the ice~Ic structure.  The ice~Ih distribution exhibits two peaks at approximately $c_{ij} =-1$ and $c_{ij} =-0.1$, indicating the presence of three eclipsed, and one ``staggered" hydrogen bond for each molecule in this crystal.

Following Refs.~\cite{rss,isv2011}, we define that a crystalline bond exists between molecule $i$ and one of its four nearest neighbours $j$ when $c_{ij} < -0.87$.  A crystalline molecule is defined as one that has three or more crystalline bonds.  Fig.~\ref{bonds} demonstrates that this definition for a crystalline molecule is unlikely to be met by molecules in the liquid phase.  Conversely, this definition is unlikely to misidentify molecules in either crystal phase as liquid-like.  

As discussed in Refs.~\cite{rss,isv2011}, this definition permits the formation of both the ice~Ic and Ih structures, since both have at least three eclipsed bonds.  As a consequence, the umbrella sampling procedure that we will apply to the liquid state will not specifically select for one of these crystal phases to the exclusion of the other.  At the same time, the possibility exists that this approach favors the ice~Ic structure over that of ice~Ih, since only three of the four eclipsed bonds of an ice~Ic molecule need to be detected, while all three of the eclipsed bonds of an ice~Ih molecule must be detected.  In the study of a patchy colloid model~\cite{isv2011}, the influence of this bias to form ice~Ic was indeed observed, although the height of the nucleation barrier was not affected.

Having identified all the crystalline molecules in the system, we can define crystalline clusters.  If a crystalline molecule $j$ is one of the four nearest neighbours of a crystalline molecule $i$, then $i$ and $j$ are assigned to the same crystalline cluster.  The size $n$ of a crystalline cluster is the total number of crystalline molecules contiguously connected in this way.  We then find $\nmax$, the size of the largest crystalline cluster in the system.  We use $\nmax$ as the order parameter to describe the progression of the system from the liquid to the crystalline phase. 

\section{Umbrella sampling}

\begin{figure}\bigskip
\centerline{\includegraphics[scale=0.35]{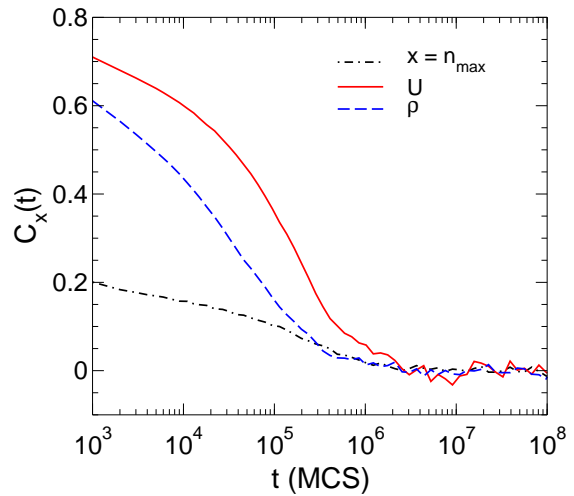}}
\caption{Time dependence of the autocorrelation functions $C_x(t)$ for $x=\nmax$, $U$, and $\rho$, as obtained from the series A run with $\rho^\ast=0.80$~g/cm$^3$ and $\nmax^\ast=19$.  This the slowest relaxing run of series A.}
\label{acf-compare}
\end{figure}

\begin{figure*}\bigskip
\centerline{\includegraphics[scale=0.6]{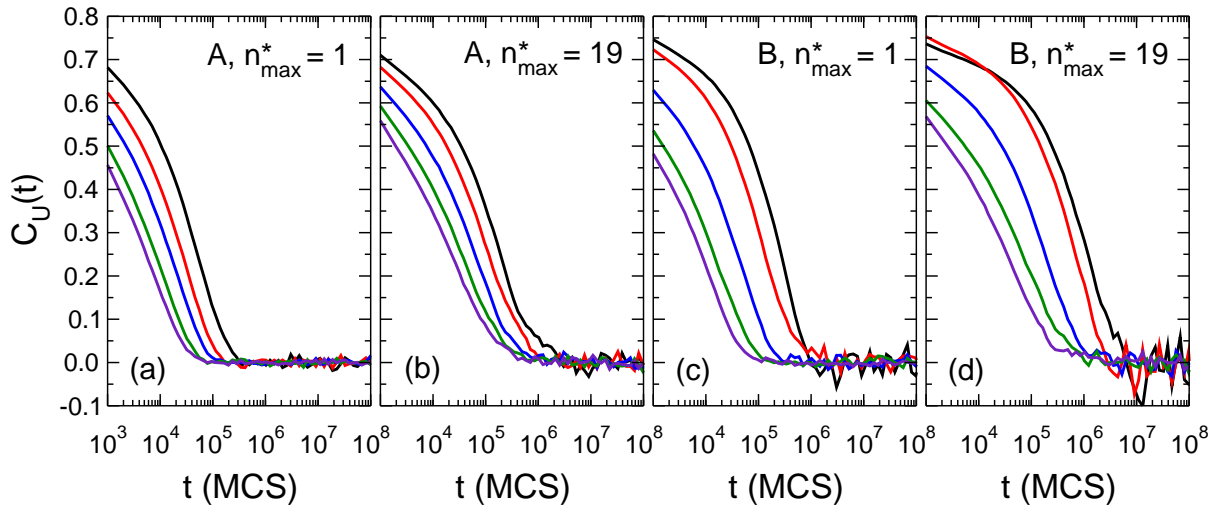}}
\caption{Time dependence of the autocorrelation function $C_U(t)$ for series A and B runs having $\nmax^\ast=1$ and $19$.  In each panel, the curves are for $\rho^\ast=0.8$, $0.9$, $1.0$, $1.1$, and $1.2$~g/cm$^3$, from most slowly decaying to most rapidly decaying.}
\label{acf-all}
\end{figure*}

\begin{figure}
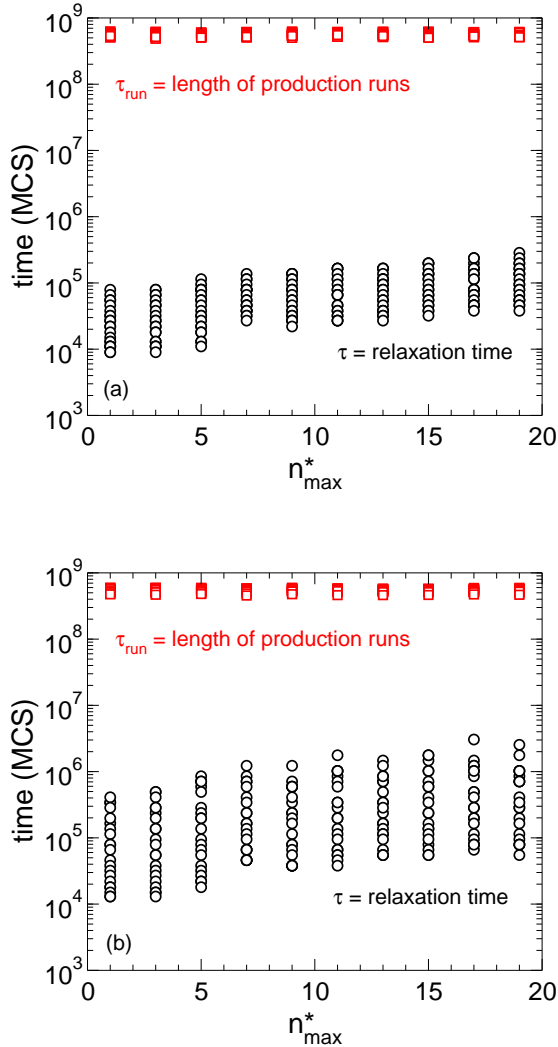
\bigskip
\centerline{\includegraphics[scale=0.35]{fig-7a.eps}}
\bigskip\bigskip
\centerline{\includegraphics[scale=0.35]{fig-7b.eps}}
\caption{Comparion of $\tau$ (circles) and $\tau_{\rm run}$ (squares) as a function of $\nmax^\ast$ for all the runs in (a) series A, and (b) series B.}
\label{times}
\end{figure}

\begin{figure}\bigskip
\centerline{\includegraphics[scale=0.35]{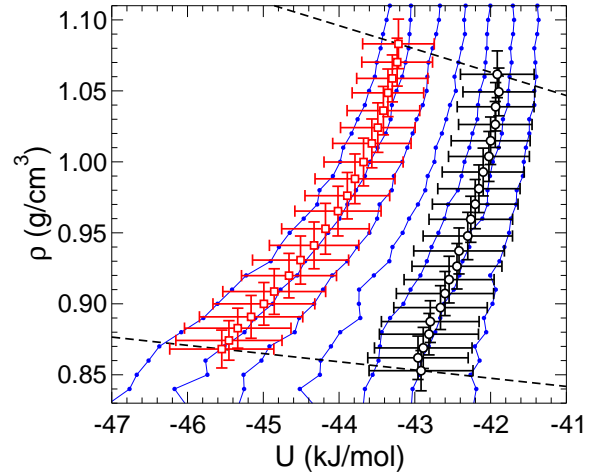}}
\caption{
Small data points connected by lines are equilibrium isotherms of $\rho$ versus $U$ as obtained from previous MD simulations~\cite{denmin}.  From left to right, the isotherms are for $T=275$ to $305$~K, in steps of $5$~K.  
Also shown are mean values of $U$ and $\rho$ at $n^*=1$ for each value of $\rho^*$ in series A (circles) and series B (squares).  The points at the lowest $\rho$ are for $\rho^*=0.80$~g/cm$^3$, and those at the highest $\rho$ are for $\rho^*=1.20$~g/cm$^3$.  The spread in the values of $U$ and $\rho$ sampled at each value of $\rho^*$ are indicated respectively by the
horizontal and vertical bars through each point.  These bars extend one standard deviation above and below the mean, for both $U$ and $\rho$.  
The lower (higher) dashed line connects the $\rho^*=0.80$~g/cm$^3$ ($\rho^*=1.20$~g/cm$^3$) points for series A and B.
The polygon in Fig.~\ref{PD} encloses all the ($T$,\,$P$) values corresponding to the data points on the equilibrium isotherms that lie between the two dashed lines.
}
\label{sampling}
\end{figure}

\begin{figure}\bigskip
\centerline{\includegraphics[scale=0.35]{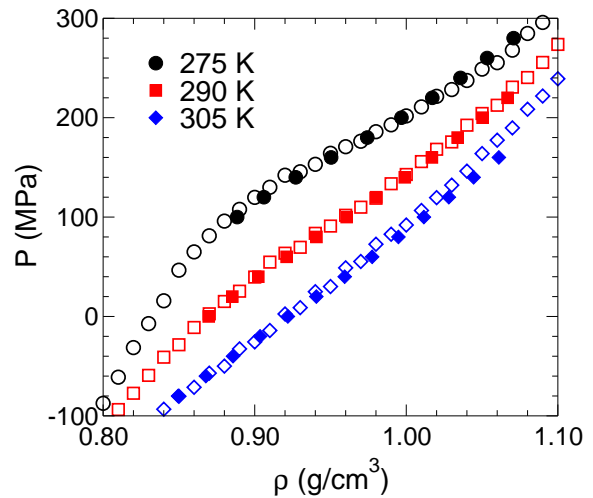}}
\caption{Comparison of isotherms of $P$ as a function of $\rho$ evaluated from the present data set using the MBAR method (filled symbols), and isotherms obtained from previous MD simulations (open symbols)~\cite{denmin}.  To facilitate comparison of the curves, the data for $T=290$~K have been shifted upward by $40$~MPa, and for $T=275$~K by $80$~MPa.}
\label{Prho}
\end{figure}

\begin{figure*}
\centerline{\includegraphics[scale=0.5]{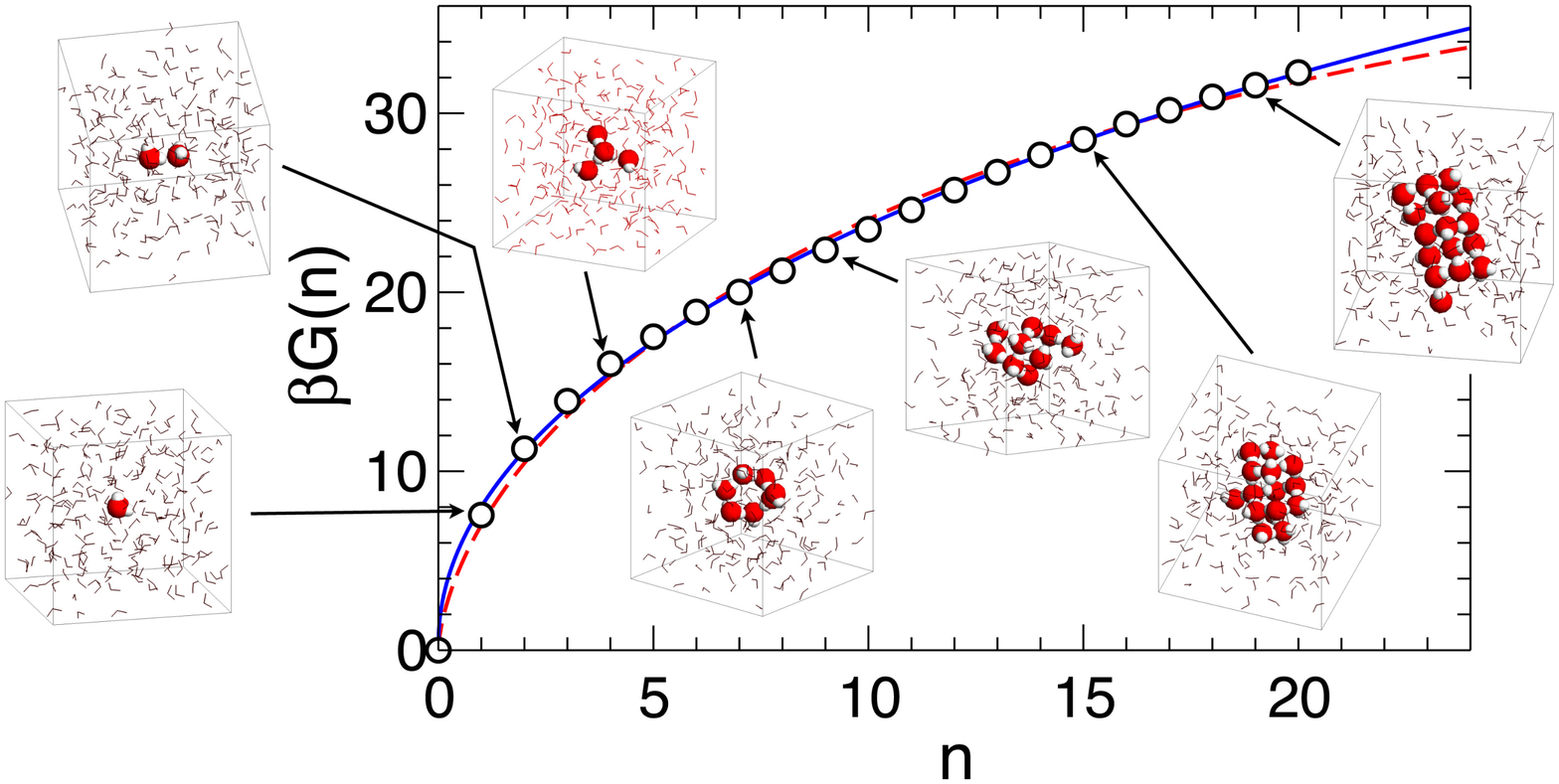}}
\caption{$G(n)$ at $T=290$~K and $P=0$~MPa (open circles), compared to fits to the data of Eq.~\ref{G1} (dashed line) and Eq.~\ref{G2} (solid line).  Also shown are sample configurations in which the molecules belonging to the largest crystalline cluster are rendered as solid spheres (red for oxygen and white for hydrogen), and where all other molecules are rendered using a stick model.  All molecules have been shifted so that the centre of mass of the largest crystalline cluster coincides with the centre of the simulation cell.  For the configurations at $n=9$, $15$, and $19$, the system has been rotated to highlight the similarity to the ice~Ic structure shown in Fig.~\ref{ices}.}
\label{bigG}
\end{figure*}

We carry out umbrella sampling MC simulations in the constant-$(N,P,T)$ ensemble.  
Our approach is similar to that used in Refs.~\cite{lim,liu,poole2013}, except that we use $\nmax$ rather than $Q_6$ as a structural order parameter.
To implement umbrella sampling, we add a biasing potential,
\begin{equation}
U_{\rm B}=k_1(\rho-\rho^\ast)^2+k_2(\nmax-\nmax^\ast)^2,
\label{umbrella}
\end{equation}
to the system potential energy $U$.  The effect of $U_{\rm B}$ is to constrain a given simulation to sample configurations in the vicinity of chosen values of the order parameters $\rho=\rho^\ast$ and $\nmax=\nmax^\ast$.  In all our simulations, we fix $N=216$, $k_1=1000kT$~(cm$^3$/g)$^2$, and $k_2=0.5kT$.

Trial configurations for each Monte Carlo step (MCS) are generated as follows:  First, we carry out an unbiased (i.e. $U_{\rm B}=0$) constant-$(N,P,T)$ MC ``sweep", in which each we attempt (on average) $N-1$ rototranslational moves, and one change of the system volume.  The maximum size of the attempted rototranslational and volume changes are chosen to give MC acceptance ratios in the range 25-40\%.  Next, the change in the biasing potential $U_{\rm B}$ is evaluated for the trial configuration resulting from the unbiased MC sweep, relative to the system configuration at the beginning of the sweep, to determine the acceptance or rejection of the trial configuration.  This completes one MCS, and the procedure is then repeated.

We next identify the $(T,P)$ state points near the Widom line at which to conduct our runs.  As a proxy for the Widom line, we use the line of compressibility maxima (LCM) shown in Fig.~\ref{PD}, as determined from previous MD simulations of the ST2 model~\cite{denmin}.  We select two state points lying on the LCM at relatively high $T$, in order to study conditions at which the relaxation time of the liquid will be relatively short.  In the following, ``series A" denotes the set of runs conducted at  $T=300~{\rm K}$ and $P=-25~{\rm MPa}$, and ``series B" denotes the set of runs conducted at  $T=280~{\rm K}$ and $P=40~{\rm MPa}$.  
The state points in the $(T,P)$ plane corresponding to series A and B are identified in Fig.~\ref{PD}.  For both series, we carry out 210 independently initialized umbrella sampling runs at equally spaced values of $\rho^\ast$ from 0.80 to 1.20~g/cm$^3$ separated by 0.02~g/cm$^3$; and equally spaced values of $\nmax^\ast$ from 1 to 19 separated by 2. 

To initialize our umbrella sampling runs, we first conduct 210 unbiased MC runs at state point A starting from independent configurations harvested from a high-temperature run conducted at $T=600$~K and $P=25$~MPa.  We separately equilibrate all of these unbiased runs at state point A.  Then the umbrella sampling potential $U_B$ is imposed, and the time evolution continued in order to bring the biased runs into equilibrium.  In all cases we find that a crystalline cluster of size $\nmax$ close to $\nmax^\ast$ develops during each run when initiated from the liquid state.  After all of the runs in series A attain equilibrium, we use a configuration from each to initiate the corresponding umbrella sampling run for series B by changing the imposed $T$ and $P$.  These series B runs are then separately evolved to equilibrium.  From each run, we record a time series of instantaneous values for $\rho$, $\nmax$, $U$, and for ${\cal N}(n)$ for $n=0$ to $20$.

Our runs are carried out for between $4.5\times 10^8$ and $6.3\times 10^8$ MCS.  
Using the second half of each run, we compute the autocorrelation function $C_x(t)$
from the time series for each choice of $x=\rho$, $\nmax$, or $U$.   As illustrated in Fig.~\ref{acf-compare}, we find in almost all cases that $U$ is the most slowly relaxing of these three quantities.  As shown in Fig.~\ref{acf-all}, in all cases $C_U(t)$ decays to zero on a time scale that is short compared to the lengths of our runs, confirming that our runs are well equilibrated.  To characterize the relaxation time of our umbrella sampling runs, we choose the time $\tau$ such that $C_U(\tau)=0.2$.  As shown in Fig.~\ref{times}, we find values of $\tau$ ranging between $9\times 10^3$ and $3\times 10^6$~MCS.  To account for equilibration, we then discard the results for $t<\tau_e$ of each run, where $\tau_e$ is chosen to be at least $20 \tau$.  The resulting length $\tau_{\rm run}$ of each production run that is used in our analysis is shown in Fig.~\ref{times}, compared to the corresponding value of $\tau$.  In terms of $\tau$, the lengths of our production runs range between 173$\tau$ and 57000$\tau$.

To estimate observables and their associated error, we use the multistage Bennet acceptance ratio (MBAR) method~\cite{mbar}.  The MBAR method takes as input the time series of the order parameters ($\rho$ and $\nmax$) and the system potential energy $U$, reweights the statistics obtained from each run to remove the effect of the biasing potential, and produces an optimal estimate of the desired observable at a specified value of $T$ and $P$.  Similar to histogram reweighting methods~\cite{ferren,wham}, the MBAR method also facilitates reweighting the configurations sampled during our runs with respect to $T$ and/or $P$, allowing the statistics from different state points to be combined to produce an estimate of a desired observable at $(T,P)$ state points that lie near the conditions at which we carry out our simulations.

For the purpose of estimating average values and their error using the MBAR method, we wish to consider only those configurations from our runs that are statistically independent.  We assume that statistically independent configurations are separated by $\tau$ or $\tau_{\rm run}/1000$, whichever is larger.  All other configurations are ignored in our analysis.  For the $G(n)$ data presented in the figures that follow, we find that one standard deviation of error is always comparable to or smaller than the symbol size.

In order to determine the range of $T$ and $P$ within which we can reliably use reweighting to estimate the average values of observables from our data set, we must examine the range of $\rho$ and $U$ sampled by our umbrella sampling procedure.  That is, we can only evaluate observables at values of $P$ where our runs have adequately sampled microstates having values of $\rho$ that are typical of equilibrium at that value of $P$.  Similarly, we can only evaluate observables at values of $T$ where our runs have adequately sampled microstates having values of $U$ that are typical of equilibrium at that value of $T$.  To proceed, we first identify in Fig.~\ref{sampling} the mean values of $U$ and $\rho$ for liquid states in the vicinity of state points A and B.  Fig.~\ref{sampling} shows equilibrium isotherms of $U$ versus $\rho$ obtained from previous MD simulations carried out at constant $(N,V,T)$, for $T=275$ to $305$~K~\cite{denmin}.  We then compare these equilibrium isotherms with the values of $U$ and $\rho$ of the microstates sampled in our $n^*=1$ runs, since these are the runs most representative of the equilibrium liquid, in which crystal nuclei are either absent or very small.  Fig.~\ref{sampling} shows the mean values of $U$ and $\rho$ sampled in our $n^*=1$ runs for all values of $\rho^*$, as well as the distribution of sampled values as characterized by the standard deviation of the set of $U$ and $\rho$ values sampled during each run.  

The dashed lines in Fig.~\ref{sampling} connect the two highest and the two lowest density points from series A and B.  We use these lines to define the limits of adequate sampling with respect to $\rho$ along each of the equilibrium isotherms.  This in turn defines the range of $P$ within which we evaluate observables.  That is, for a given $T$ we only consider the range of $P$ that corresponds to the range of $\rho$ found between these two dashed lines.  This criterion determines the upper and lower boundaries of the polygon shown in Fig.~\ref{PD}.

To delimit the range of $T$ examined, our criterion is that the equilibrium isotherms in Fig.~\ref{sampling} should fall within one standard deviation of the mean values of $U$ sampled in our umbrella sampling runs.  This criterion is satisfied between $275$ and $305$~K, except for $T$ in the vicinity of $290$~K.  However, the $290$~K equilibrium isotherm lies between the series A and B data sets, and is also within two standard deviations of both.  Consequently, the statistics of both series contribute to the evaluation of observables near $290$~K.  As we will see below, we confirm that the statistics accumulated in our umbrella sampling runs are indeed sufficient to compute observables throughout the range of $T$ from $275$ to $305$~K, thus justifying the high and low $T$ boundaries of the polygon shown in Fig.~\ref{PD}.

To test our ability to evaluate average values from our data set over the range of $T$ and $P$ indicated by the polygon in Fig.~\ref{PD}, we show in Fig.~\ref{Prho} isotherms of $P$ versus $\rho$, for $T$ and $P$ spanning the width and height of the polygon.  Fig.~\ref{Prho} demonstrates good agreement between the isotherms calculated by using the MBAR method to find the average value of $\rho$ at a given $T$ and $P$ from our combined series A and B data sets, and the isotherms obtained from previous MD simulations~\cite{denmin}.  

\section{Results}

To find $G(n)$ throughout the $(T, P)$ range within the polygon of Fig.~\ref{PD}, we combine the statistics from all runs in series A and B.  We use the MBAR method to compute the average value of ${\cal N}(n)$ for $n=0$ to $20$, on a grid of $(T, P)$ points within the polygon, where each point is separated by $5$~K in $T$ and $20$~MPa in $P$.  $G(n)$ at each state point is then found from ${\cal N}(n)$ using Eq.~\ref{G}.

A representative result is given in Fig.~\ref{bigG}, which shows $G(n)$ found at $T=290$~K and $P=0$~MPa, a point very close to the LCM.  Fig.~\ref{bigG} shows that there is a large free energy cost (up to $30kT$) for the formation of small ice nuclei  in the vicinity the $K_T$ maximum, demonstrating that the thermodynamic anomalies associated with the Widom line for ST2 water occur in a well defined metastable liquid, for which crystal formation can only occur by surmounting a substantial free energy barrier.

\begin{figure}\bigskip
\centerline{\includegraphics[scale=0.35]{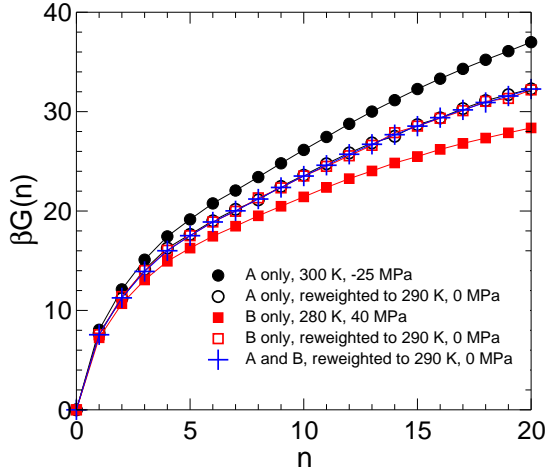}}
\caption{Test of the results for $G(n)$ obtained using the MBAR method with reweighting to $T=290$~K and $P=0$~MPa, using three different approaches:  (i) using reweighted data from series A only (open circles); (ii) using reweighted data from series B only (open squares); and using reweighted and combined data from both series A and B (plus signs).  For comparison, we also show $G(n)$ curves (filled symbols) obtained from series A and B without reweighting with respect to $T$ or $P$.}
\label{testMBAR}
\end{figure}

\begin{figure}
\centerline{\includegraphics[scale=0.35]{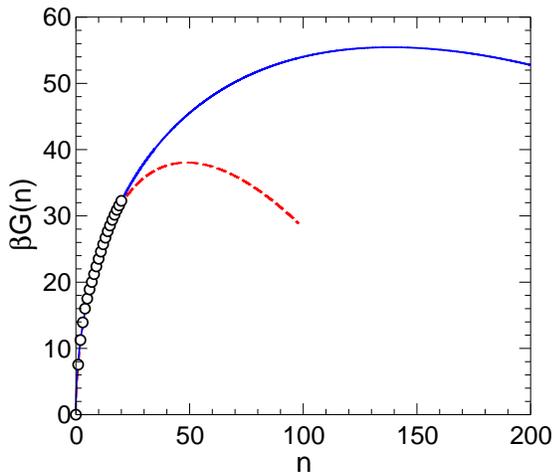}}
\caption{$G(n)$ at $T=290$~K and $P=0$~MPa (open circles), compared to fits to the data of Eq.~\ref{G1} (dashed line) and Eq.~\ref{G2} (solid line).}
\label{wideG}
\end{figure}

\begin{figure}
\centerline{\includegraphics[scale=0.35]{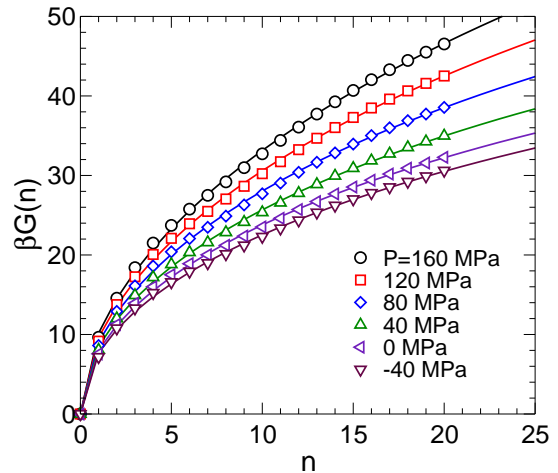}}
\caption{$G(n)$ at $T=290$~K over a range of $P$.  Solid lines are fits of Eq.~\ref{G2}.}
\label{G290}
\end{figure}

In order to confirm the result presented in Fig.~\ref{bigG}, and to test for systematic errors in our implementation of the MBAR method (including reweighting), we conduct an additional test.  As shown in Fig.~\ref{testMBAR}, we compare the $G(n)$ curve from Fig.~\ref{bigG} (obtained using data from both series A and B) with the result obtained from series A alone, and from series B alone.  All three curves are within error of one another.  This agreement confirms that the statistics gathered from series A and B have independently converged to equilibrium, and that the reweighting of data gathered at either state point A or B gives consistent results at $T$ and $P$ intermediate between A and B.

Visual inspection of the structure of the larger ice nuclei (in the range of $n=15$ to $20$) suggests that most of the nuclei formed in our simulations have the structure of ice~Ic, as illustrated in Fig.~\ref{bigG}.  We have not found any examples of ice~Ih nuclei.  As discussed in Section~II, this result may be a consequence of the order parameter used in our umbrella sampling procedure, or it may indicate that for the ST2 model the free energy required to form small ice~Ih nuclei is higher than for ice~Ic, under the conditions studied here.  Following the approach of Ref.~\cite{isv2011}, this question could be resolved in future work by using modified order parameters that enforce the formation of only ice~Ic or ice~Ih.

To test if our results may be used to estimate the height of the nucleation barrier $G^*$ and the size of the critical nucleus $n^*$, we use our data for $G(n)$ to obtain a fit of an equation for $G(n)$ the form of which is predicted by classical nucleation theory:
\begin{equation}
G(n)=-an+bn^{2/3},
\label{G1}
\end{equation}
where $a$ and $b$ are fit parameters~\cite{pdbook}.  We also fit a modified equation that includes an additional term and fit parameter $c$:
\begin{equation}
G(n)=-an+bn^{2/3}+cn^{1/3}.
\label{G2}
\end{equation}
A term of the form of that added in Eq.~\ref{G2} has been used to account for the curvature of the nucleus-liquid interface for simple liquids~\cite{ls}, as well as to model a core-shell morphology of the nucleus in ice nucleation~\cite{russo2014}.

The fits of Eq.~\ref{G1} and \ref{G2} obtained for $G(n)$ at $T=290$~K and $P=0$~MPa are shown in Figs.~\ref{bigG} and Fig.~\ref{wideG}.  While both fitting functions do a reasonable job of representing the data in the range of small $n$, they give widely different estimates for $G^*$ and $n^*$.  Similar results to those shown in Fig.~\ref{wideG} are obtained for state points throughout the $(T,\, P)$ region in which we compute $G(n)$.  Given the large degree of extrapolation required to estimate $G^*$ and $n^*$ from our data, we do not attempt to analyze these quantities further here.

The variation of $G(n)$ with $P$ at $T=290$~K is shown in Fig.~\ref{G290}.  For a given $n$, the value of $G(n)$ increases with $P$.  This is in line with expectation.  At $T=280$~K and $P=40$~MPa, the density of ice~Ic and Ih is $0.862$~g/cm$^3$, whereas the density of the liquid is $0.917$~g/cm$^3$.  At higher $P$, the difference in density between the liquid and crystal increases.  Hence the local structure of the liquid is more tetrahedral and ice-like at lower $P$, where the densities of the liquid and crystal are closer.  As $P$ increases, the local tetrahedral network of the liquid is progressively disrupted, and a more significant local restructuring is required to create an ice-like cluster.  From this point of view, we should expect the free energy cost to create ice nuclei to increase with $P$ in the range of pressure examined here.

\begin{figure}
\bigskip
\centerline{\includegraphics[scale=0.35]{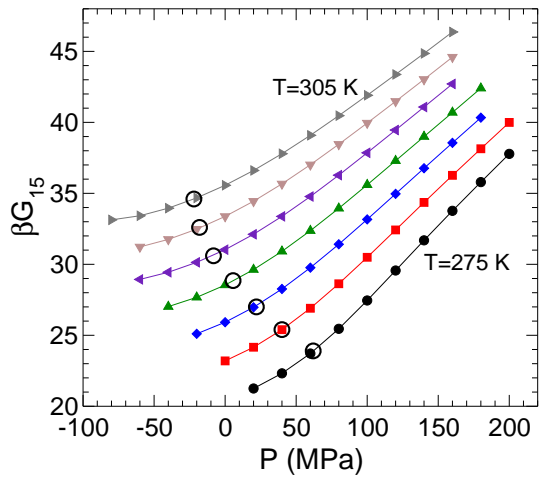}}
\caption{Isotherms of $G_{15}$ as a function of $P$.  Isotherms are $5$~K apart.  The open circles locate the pressure of the $K_T$ maximum along each isotherm.}
\label{G15T}
\end{figure}

\begin{figure}
\bigskip
\centerline{\includegraphics[scale=0.35]{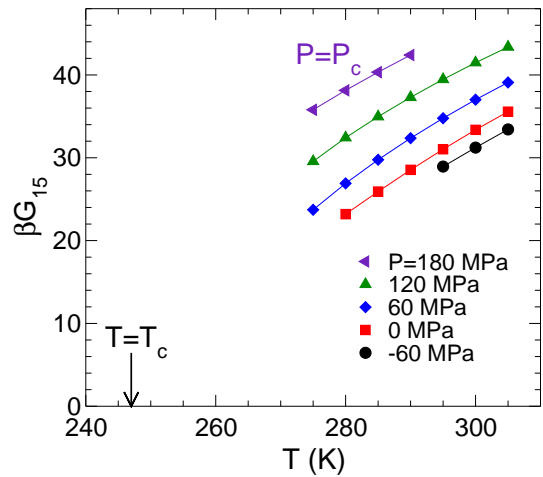}}
\caption{Isobars of $G_{15}$ as a function of $T$.  $T_c$ and $P_c$ are respectively the estimated $T$ and $P$ of the critical point of the LLPT.  
}
\label{G15P}
\end{figure}

To characterize the behaviour of $G(n)$ over the entire $(T,\, P)$ range studied here, we focus on the value of $G_{15}=G(n=15)$.  Isotherms of $G_{15}$ are shown in Fig.~\ref{G15T}, and isobars in Fig.~\ref{G15P}.  The two dimensional function $G_{15}(T,P)$ is represented in Fig.~\ref{PD} as a contour plot.  We find that over the entire region examined, $G_{15}$ remains large relative to $kT$, ranging from $22kT$ to $46kT$.  $G_{15}$ monotonically increases with $P$ along isotherms (Fig.~\ref{G15T}), and monotonically increases with $T$ along isobars (Fig.~\ref{G15P}).

As shown in Fig.~\ref{G15T}, we note that the isotherms of $G_{15}$ are approximately linear in $P$ at high $P$, but then start to curve as $P$ passes through the vicinity of the $K_T$ maximum.  Indeed, there is reason to predict that these isotherms will pass through a minimum at negative $P$ beyond the range of our current data.  The density of the liquid and crystal are approaching one another as $P$ decreases, and can be expected to intersect in the negative pressure region.  Near this intersection, the free energy cost to create ice nuclei of a given size should be lower than at higher or lower $P$, since under these conditions an ice embryo can form without a density fluctuation.  This picture is also consistent with the requirement (imposed by the Clapeyron equation) that the ice-liquid coexistence line must pass through a maximum $T$ as $P$ decreases if the liquid and crystal densities intersect.  Such a melting-line maximum has been observed for other water models; see e.g. Ref.~\cite{vegamelt}.  For a supercooled liquid near a melting-line maximum, the ice nucleation barrier must have a minimum as a function of $P$ along an isotherm because the coexistence line (where the nucleation barrier diverges) is crossed by both increasing or decreasing $P$.

Consistent with the above considerations, the onset of curvature in the isotherms of $G_{15}$ as $P$ decreases through that of the $K_T$ maximum reflects the change in liquid structure that occurs as the Widom line is crossed.  As $P$ decreases below that of the $K_T$ maximum, the liquid enters the region where it becomes increasingly similar to the LDL phase, characterized by a high degree of local tetrahedral structure and a density approaching that of ice.  Our data show that the onset of this crossover in the liquid phase is also reflected in the behaviour of the free energy of formation of small ice nuclei.

The isobars of $G_{15}$ presented in Fig.~\ref{G15P} show that $G_{15}$ decreases with $T$ at all $P$ studied here.  Fig.~\ref{G15P} also identifies the isobar corresponding to the estimated critical pressure $P_c=180$~MPa of the LLPT for the ST2 model, as well as the value of the critical temperature $T_c=247$~K~\cite{cuth}.  In the case that the supercooled liquid becomes unstable to ice formation, $G^*$ and $n^*$ should approach zero, in which case $G_{15}$ should also approach zero.  Our isobars of $G_{15}$ are too far away from $T_c$ to justify a quantified extrapolation.  Nonetheless, our data suggests that future work to extend these curves to lower $T$ would provide useful information.  On the one hand, the slope of our isobar for $P=P_c$ suggests that a non-zero ice nucleation barrier exists in the vicinity of the critical point of the LLPT, a result that is consistent with several recent free energy simulation studies of ST2 water~\cite{liu,poole2013,rsc,palmer2014}.  On the other, our isobars for $P<P_c$ raise the possibility that the ice nucleation barrier may become low or even approach zero in the region of the phase diagram below the coexistence line of the LLPT, in the stability field of the LDL phase.  If such a stability limit for the LDL phase of ST2 water exists, clarifying its location would be valuable for understanding this phase of supercooled water, and for elucidating the influence of the LLPT on the ice nucleation process.

\section{Discussion}

In summary, we have quantified the free energy of formation of small ice nuclei in the ST2 model of water.  For the range of $T$ and $P$ studied here, we find that these small nuclei have the local structure of ice Ic, and that there is a significant free energy cost to form these ice nuclei in the vicinity of the Widom line.  This result excludes the possibility that the liquid state of the ST2 model is unstable to ice formation in this region of the phase diagram, and thus confirms that the thermodynamic anomalies previously identified in this region (e.g. the $K_T$ maximum) occur in a well-defined metastable liquid state that is separated from the crystal by a substantial free energy barrier.  A similar conclusion was recently drawn for the case of the TIP4P/2005 model of water, though by different means~\cite{vega,sanz2014}.

From a methodological standpoint, our work illustrates that two-dimensional umbrella sampling can be exploited to quantify nucleation phenomena over a continuous range of $T$ and $P$, and is a valuable approach when seeking to correlate thermodynamic behaviour and nucleation processes in simulations of supercooled liquids.

For the specific case of supercooled water, our results show that we can detect the crossover from the HDL-like liquid to the LDL-like liquid by examining the nucleation behaviour of ice from the liquid phase.  Since this crossover influences the free energy of formation of ice nuclei, we can also expect it to influence the nucleation rate of ice.  This observation is significant because the principal challenge for the direct experimental characterization of deeply supercooled water is the increasingly rapid conversion of the metastable liquid to ice as $T$ decreases.  However, the ice nucleation process is strongly influenced by the nature of the metastable liquid in which the process begins, and our results demonstrate that ice nucleation can be used as a ``probe" to reveal the properties of the underlying supercooled liquid.  Our work raises the possibility that an examination of ice nucleation rates in deeply supercooled water might be useful as a means to locate and study the thermodynamic anomalies of the supercooled liquid, including perhaps the LLPT itself.  

\begin{acknowledgements}
CRCB, RKB, ISV, and PHP thank NSERC for support.  Computational resources were provided by ACEnet.  
We thank
S.~McGibbon-Gardner,
S.~Morris, 
and O.~Zavalov
for useful discussions.
\end{acknowledgements}

\end{document}